\documentclass{PoS}

\usepackage[T1]{fontenc}
\usepackage{amssymb}
\usepackage{times}
\usepackage{mathptmx}
\usepackage{graphicx}
\usepackage{sidecap}
\usepackage{wrapfig}
\usepackage{xspace}

\newcommand{\gev}{\,{\rm GeV}\xspace}

\newcommand{\PYTHIA}{{\sc Pythia}\xspace}
\newcommand{\HERWIG}{{\sc Herwig}\xspace}

\newcommand{\CASCADE}{{\sc Cascade}\xspace}
\newcommand{\MCATNLO}{{\sc MC@NLO}\xspace}

\author{Zlatka Staykova on the behalf of H1 Collaboration}
\title{Photoproduction of $D^*$ and Jets at H1}
\abstract{Photoproduction events containing a charmed meson $D^{*\pm}$ and
  two jets were investigated with the H1 detector using the HERA II
  data sample. The $D^*$ meson was reconstructed in the decay
  channel, $D^{*\pm}\rightarrow D^0\pi^\pm\rightarrow K^\mp\pi^\pm\pi^\pm$. Jets were
  reconstructed using the inclusive $k_t$ algorithm and were selected
  if they have transverse momenta $p_t(\textrm{jet})>3.5\gev$. One of the jets was associated with the $D^*$ meson itself,
  such that the jet originating from the parent charmed quark as the
  meson can be tagged. The phase space of the measurement is limited within central
  rapidity for the $D^*$ meson and the $D^*_\textrm{jet}$,
  $|\eta|<1.5$ while the second jet was measured within, $-1.5<\eta<2.9$. Single
  differential cross sections and double differential distributions
  were measured and compared to Leading Order Monte Carlo (MC) event generators, \PYTHIA and \CASCADE
  and with the Next--to--Leading order MC generator \MCATNLO}

\FullConference{XVII International Workshop on Deep-Inelastic
  Scattering and Related Subjects,\\ Convitto della Calza, Firenze,
  Italy, \\April 19 -23, 2010}
\ShortTitle{$D^*$ and Jets at H1 in $\gamma p$}

\begin{document}
\maketitle

\section{Introduction and Motivation}
\begin{wrapfigure}{r}{0.4\textwidth}
  \centering
  \includegraphics[width=0.4\textwidth]{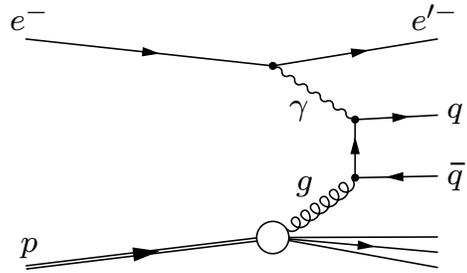}\\
%  \vspace{0.6cm}
%  \includegraphics[width=0.2\textwidth]{../../graphics/DIS10/resolved.eps}
  \caption{A Feynman diagram of the boson gluon fusion process.}
  \label{fig:bgf}
  
\end{wrapfigure}

The dominant production process of charmed quarks in DIS is the boson
gluon fusion process (BGF) (Fig. \ref{fig:bgf}), where a gluon splits into
quark anti--quark pair and then interacts with a virtual photon. Such
charm production is highly sensitive to the gluon
content of the proton. Moreover charm mass $m_c\simeq 1.5\gev$
provides the hard scale for perturbative calculations in the regime of
photoproduction where the photon virtuality approaches zero. In
summary charm quarks in photoproduction provide good testing ground for
pQCD calculations. 

Higher order processes are often approximated by parton showers, where
the successive emitted partons from the proton side are ordered in a given kinematic
quantity. Two major approaches are used, the DGLAP and
the CCFM evolution equations. The basic difference between them is the quantity used for the
ordering, in the case of DGLAP the transverse momentum of the partons
is used, while in CCFM the emission angle with respect to the incoming
gluon.

%However higher order gluon emissions are still possible. The treatment
%of these emissions is normally done with the so-called parton shower
%approaches. In this concept the parton from the proton side emits
%ordered successive partons ordered in a kinematic quantities. The most
%common approach is the DGLAP approach where the partons are ordered in
%virtuality (or transverse momentum) limited from above by the photon
%virtuality $Q^2$ (or the average transverse mass of the quark
%pair). In the construction of this approach the highest transverse momentum is
%the one of the quarks and emission of partons with relatively large
%transverse momentum below is not likely. However, in the regime of
%photoproduction the virtual photon is close to its mass shell and
%fluctuations to quark anti--quark pairs is very possible. This regime is
%called resolved photoproduction. Resolved photoproduction processes
%shift the hard interaction towards the proton (figure \ref{fig:bgf}, bottom) such that high
%transverse momentum emissions closer to the proton side is possible.

In photoproduction the photon is quasi--real can split into partons. These events are called \emph{resolved photoproduction} and
are treated in terms of photon parton density functions (pdf) and
evolution equations can be also applied to the photons. This is of
particular need for the DGLAP based models \cite{Jung:1999eb}. The
process when the photon interacts as a pointlike object is called
\emph{direct photoproduction}.

A previous measurement \cite{Aktas:2006ry} of photoproduction of $D^*$ mesons and two
jets at H1, was found to be highly sensitive to
the two different parton shower approaches. In the present measurement
a larger data sample is used and the phase space
was extended towards larger rapidity for the second jet.

{\bf Monte Carlo Models}\\
The presented measurement was compared to two leading order Monte
Carlo (MC) programs, \PYTHIA \cite{Sjostrand:2001yu} and
\CASCADE \cite{Jung:2001hx}. In \PYTHIA the parton showers are implemented according to
the DGLAP evolution equations. It generates also resolved photon
processes. Two different modes of \PYTHIA were used: in the
first case the matrix elements were calculated explicitly for heavy
quarks, and the final prediction is a mixture of two different
samples, direct and resolved processes. For that case the SAS 2D LO
photon pdfs were used as well as the CTEQ 6M NLO proton pdfs. In the
second mode the generator calculates the matrix elements for massless quarks. In this case the
photon pdfs were GRV-G LO and the CTEQ 6L LO proton pdfs were also
used. The \CASCADE MC generated the parton showers according to the
CCFM evolution equation with the set A0 unintegrated proton pdfs. A next--to--leading order MC
generator, (MC@NLO) \cite{Toll:2010zz} was also compared to the data. It is a full NLO
matrix elements for heavy quark photoproduction matched with partons
showers. The parton showers are implemented according to the DGLAP evolution
equations, and were taken from the \HERWIG MC program and the CTEQ 6.6
proton pdfs were used. The uncertainty of the calculation was
estimated varying the factorisation and renormalisation scales.
\section{Experimental Setup, Event Selection and Event Reconstruction}
\begin{wrapfigure}{r}{0.4\textwidth}
  \centering
  \includegraphics[width=0.4\textwidth]{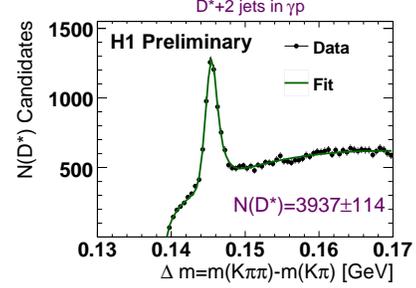}
  \caption{The mass difference $\Delta m=m(K\pi)-m(K\pi\pi_{slow})$}
  \label{fig:dm}
\end{wrapfigure}

In this measurement the HERA II data sample collected with the H1
detector \cite{Abt:1996hi} was used, corresponding to an integrated luminosity of $\mathcal
L=93.4\,{\rm pb}^{-1}$. The measurement was performed in untagged
photoproduction. The scattered electron escapes detection and the
event kinematic variables were reconstructed using the hadronic final
state (HFS). For
the online triggering of the events the fast
track trigger (FTT) \cite{Schoning:2006zc} was used. Charged particles are reconstructed
online and offline and combined into a $D^*$ meson candidates using the
mass hypothesis. The $D^*$ was reconstructed in the
channel: $D^{*\pm}\rightarrow D^0\pi^\pm\rightarrow K^\mp\pi^\pm\pi^\pm$. The $D^*$ mesons were
selected if they have transverse momentum $p_t>2.1\gev$ limited
in the central rapidity range by the central tracking device $|\eta|<1.5$. The jets were
reconstructed with the inclusive $k_t$ algorithm
\cite{Cacciari:2005hq} in the laboratory
frame in the energy recombination scheme. The $D^*$ was treated as a
leading particle, which means that the four vectors of the decay products of the meson
were replaced by the four vector of the $D^*$ itself in the HFS
definition. Such, the $D^*_\textrm{jet}$ could be identified. The
second hardest jet 
in the event is denoted as the \emph{other
jet}. The jets were selected if they have transverse momentum of
$p_t^{\textrm{jet}}>3.5\gev$ in the central rapidity range for the
$D^*_\textrm{jet}$, $|\eta(D^*_\textrm{jet})|<1.5$ to be consistent
with the $D^*$ selection. The other jet was measured in the range
$-1.5<\eta(\textrm{Other jet})<2.9$. Finally a cut on the invariant
mass of the jets $M_{\textrm{jj}}>6\gev$ was applied.

The number of signal events was determined from a fit to the $D^*$ mass difference
distribution $\Delta m=m(K\pi)-m(K\pi\pi)$, (Fig. \ref{fig:dm}). For the signal the asymmetric Crystal Ball \cite{Gaiser:1982yw}
function was used. The fit was performed in each bin of the
measurement. Within the selection about 4000 $D^*$ mesons were found. 

Various sources of systematic uncertainty were investigated leading to
total systematic uncertainty of about 10\%. The dominant sources are
the luminosity measurement, track finding and the trigger efficiency uncertainties.

\section{Results of the Measurement}
%\begin{wraptable}{r}{0.3\textwidth}
%  \centering
%  \small
%  \begin{tabular}{c}
%    \hline
%    $0.1<y_h<0.8$\\
%    $Q^2<2.\gev^2$\\
%    $2.1\gev<p_t(D^*)<12.5\gev$\\
%    $\left|\eta(D^*)\right|<1.5$\\
%    $3.5\gev<p_t(D^*_{\rm{jet}},\rm{Other~jet})<15.\gev$\\
%    $\left|\eta(D^*_{\rm{jet}})\right|<1.5$\\
%    $-1.5<\eta(\rm{Other~jet})<2.9$\\
%    $M_{jj}>6\gev$\\
%    \hline
%  \end{tabular}
%  \caption{The phase space definition}
%\label{tab:phaseSpace}
%\end{wraptable}

\begin{figure}[h]
  \centering
  \includegraphics[width=0.8\textwidth]{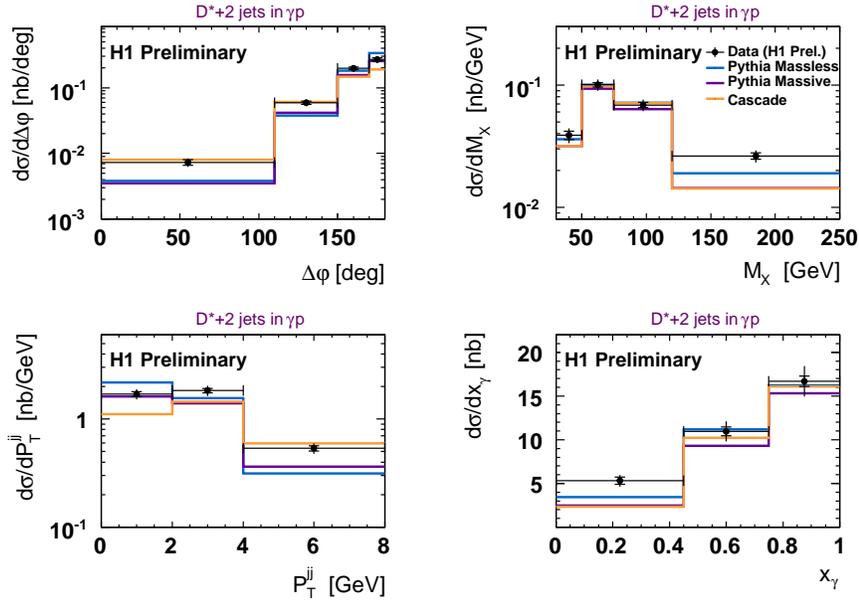}
  \caption{The differential cross sections for $D^*$ and two jet
    production in photoproduction as a function of $\Delta\varphi$
    between the jets, $M_X$, $p_t^{\textrm{jj}}$ and $x_\gamma$.}
  \label{fig:Corr}
\end{figure}

\begin{figure}[h!]
  \centering
  \includegraphics[width=0.49\textwidth]{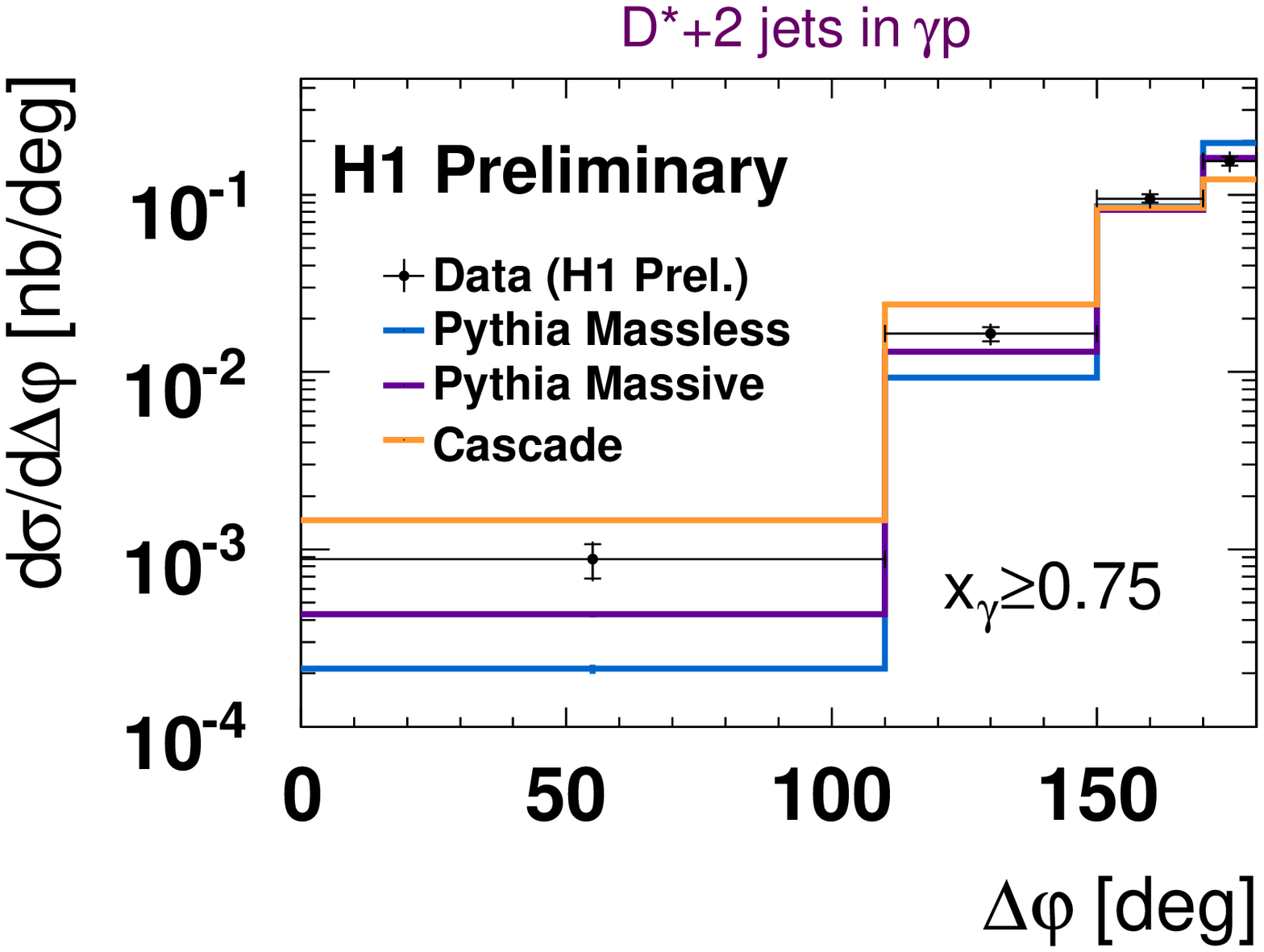}
  \includegraphics[width=0.49\textwidth]{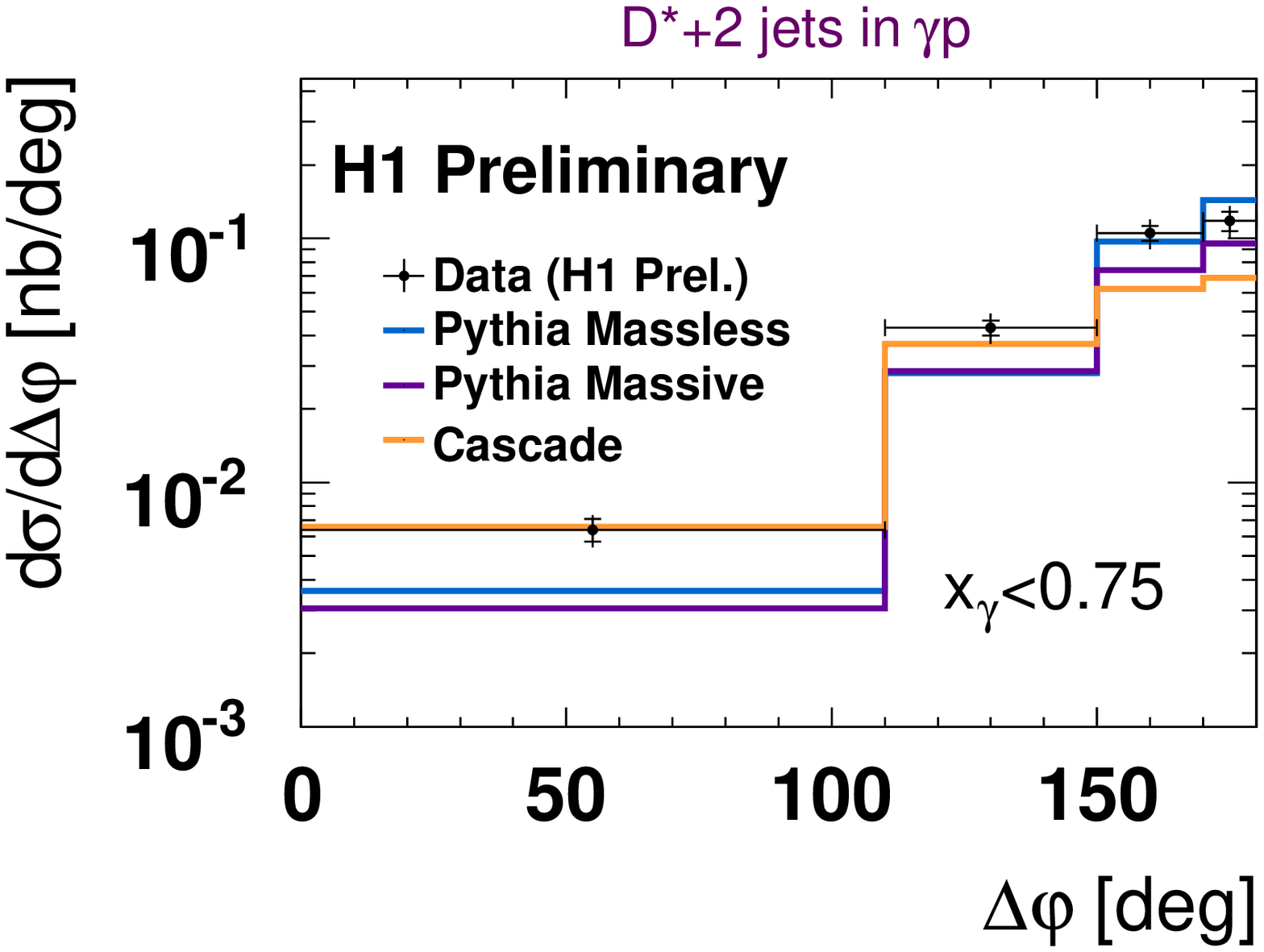}\\
  \caption{The $\Delta\varphi$ distribution for two bins of
    $x_\gamma$, compared to LO MC programs}
  \label{fig:DPhi}
\end{figure}

\begin{figure}[h1]
  \centering
  \includegraphics[width=0.483\textwidth]{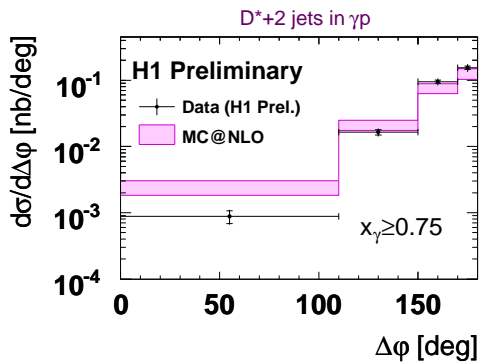}\\
  \caption{The $\Delta\varphi$ distribution for the high $x_\gamma$
    region compared to MC@NLO}
  \label{fig:MCNLO}
\end{figure}

The cross sections are measured in the phase space as summarised
above. The results are presented in Fig.\ref{fig:Corr} --
Fig.\ref{fig:MCNLO}. The cross sections are represented by 
the black markers, where the inner error bars shows the systematic
uncertainty and the outer error bars are the total uncertainty.

%\begin{wrapfigure}{r}{0.6\textwidth}
The kinematic quantities of the $D^*$ and
the jets (not shown here) are well reproduced by the MC generators. In Fig. \ref{fig:Corr} the differential cross section of
$D^*$ and two jets in photoproduction are shown as a function of the azimuthal
angle difference between the two jets $\Delta\varphi$ (top
left); the invariant mass of the remnant in the event $M_X$ (top
right); the average transverse momentum of the di--jet pair
$p_t^{\textrm{jj}}$ (bottom left) and the longitudinal momentum
fraction of the photon carried by the jets
$x_\gamma$ (bottom right). Here, the shape of the azimuthal angle
difference $\Delta\varphi$ is not well reproduced by three models. The
back--to--back region $\Delta\varphi$ close to $180^\circ$, where the jets are
balanced in $p_t$ and contribution from higher order gluon radiation
is not expected, is well modeled by \PYTHIA massive. In the region towards small
$\Delta\varphi$, further radiation is expected from momentum conservation, the data are well predicted by \CASCADE. The same
can be observed in the average transverse momentum of the jet pair
$p_t^{\textrm{jj}}$. The invariant mass of the remnant $M_X^2=(p+ \gamma -(v_1+v_2))^2$, where $v_{1/2}$ are the four vectors
of the two jets, is found to be well described by the models. The $x_\gamma$ distribution is well described
by the models with an exception of the lowest bin, where the
contributions from events containing resolved photons is dominant.

The $\Delta\varphi$ distribution was also measured for two bins of $x_\gamma$ (Fig. \ref{fig:DPhi}). For the
direct photon case, $x_\gamma\geq 0.75$, within one
sigma the distribution is fairly well modelled by \CASCADE. It can be seen that at small $\Delta\varphi$ the
model is slightly above the data. At the smallest $\Delta\varphi$ bin
both \PYTHIA models are underestimating the data by a factor of 5. In the
resolved case, $x_\gamma<0.75$, \CASCADE significantly underestimates
the data in the back--to--back region, with a difference of factor of
4 while at the tail of the distribution towards small $\Delta\varphi$
the model describes the data. The
predictions of \PYTHIA for the smallest $\Delta\varphi$ bin is
somewhat better than for the direct case. The data were also compared to
MC@NLO. It was found that for the low $x_\gamma<0.75$ (not shown here) region the
predictions are below the data but for the direct photon $x_\gamma\geq
0.75$ case the shape is very well reproduced (see Fig. \ref{fig:MCNLO}).
%\begin{wrapfigure}{l}{0.4\textwidth}
%\begin{wrapfigure}{l}{0.3\textwidth}
\section{Conclusions}
Cross sections of charm photoproduction with jets are highly sensitive
to different parton shower models. A new measurement of $D^*$ and two jets
in photoproduction with the H1 detector was presented and differential
cross sections and double differential distributions were measured with a larger
data sample and in an extended phase space than a previous measurement
by H1. Two new observables $M_X$ and $p_t^{\textrm{jj}}$ were
introduced and measured. The measured cross sections were compared with predictions
using different MC generators based on different parton shower
models. It was found that the cross sections as a function of
$\Delta\varphi$ shows very different shape for the data and the
models.

\end{document}